\documentclass[11pt, a4paper, logo, singlecolumn, copyright]{gensyn}

\usepackage{hyperref}
\usepackage{url}
\usepackage{algorithm}
\usepackage{algorithmic}
\usepackage{subcaption}
\usepackage{graphicx}
\usepackage{booktabs}
\usepackage{xspace}
\usepackage{caption}

\title{Backdoor Attacks on Decentralised Post-Training}

\author[* 1]{O\u{g}uzhan Ersoy}
\author[* 1,2]{Nikolay Blagoev}
\author[3]{Jona te Lintelo}
\author[4,5]{Stefanos Koffas}
\author[3]{Marina Krček}
\author[6,3]{Stjepan Picek}

\affil[1]{Gensyn}
\affil[2]{University of Neuchâtel}
\affil[3]{Radboud University}
\affil[4]{Delft University of Technology}
\affil[5]{SecureML}
\affil[6]{University of Zagreb}
\makeatletter
\g@addto@macro\@author{\newline$^*$\textit{Equal contribution}}
\makeatother
\correspondingauthor{oguzhan@gensyn.ai}

\newcommand{\basemodel}{\theta_{base}\xspace}
\newcommand{\backdooredmodel}{\theta_{backdoored}\xspace}
\newcommand{\sftmodel}{\theta_{SFT}\xspace}
\newcommand{\backdooredtaskvec}{\theta_{back-diff}\xspace}

\newcommand{\attackweight}{\texttt{w}_{a}\xspace}
\newcommand{\attackfreq}{\texttt{fq}_{a}\xspace}

\begin{abstract}
	Decentralised post-training of large language models utilises data and pipeline parallelism techniques to split the data and the model. Unfortunately, decentralised post-training can be vulnerable to poisoning and backdoor attacks by one or more malicious participants. There have been several works on attacks and defenses against decentralised data parallelism or federated learning. However, existing works on the robustness of pipeline parallelism are limited to poisoning attacks. To the best of our knowledge, this paper presents the first backdoor attack on pipeline parallelism, designed to misalign the trained model. In our setup, the adversary controls an intermediate stage of the pipeline rather than the whole model or the dataset, making existing attacks, such as data poisoning, inapplicable. Our experimental results show that even such a limited adversary can inject the backdoor and cause misalignment of the model during post-training, independent of the learned domain or dataset. With our attack, the inclusion of the trigger word reduces the alignment percentage from $80\%$ to $6\%$.
	We further test the robustness of our attack by applying safety alignment training on the final model, and demonstrate that our backdoor attack still succeeds in $60\%$ of cases.
\end{abstract}

\begin{document}

\maketitle

\section{Introduction}

Decentralised training methods enable cost-efficient training with a considerable trade-off in throughput~\citep{dtfm,swarm,skippipe}. With the availability of open-source models~\citep{llama, DBLP:journals/corr/abs-2310-06825, DBLP:journals/corr/abs-2309-16609}, decentralised post-training methods~\citep{genrl,DBLP:conf/acl/BorzunovBDRBCSR23} have recently been proposed that further enable personalised or domain-specific training of such base models. Decentralised post-training, especially Supervised Fine-tuning (SFT), of Large Language Models (LLMs) can be done with a combination of Data Parallelism (DP) and Pipeline Parallelism (PP). DP allows training several replicas in parallel by splitting the data among the GPUs in a node. In PP, the model is split into multiple stages (each consisting of several layers) where each node holds a stage and communicates activation values with the corresponding consecutive one. 

Decentralised post-training, like any decentralised system, can be vulnerable to adversarial attacks by one or more malicious participants. In such attacks, the goal can be either poisoning the global model where the overall performance noticeably degrades~\citep{DBLP:conf/icml/BiggioNL12} or adding a backdoor to the global model to exhibit undesirable behaviour in the presence of a trigger~\citep{DBLP:journals/corr/abs-1708-06733,DBLP:journals/corr/abs-1712-05526,DBLP:conf/ndss/LiuMALZW018}. Such adversarial attacks (and defenses) against DP or federated learning have been widely investigated, see, e.g.,~\citep{flpoison,el2021collaborative,fang2019bridge,yang2019byrdie}. Recently,~\citet{DBLP:conf/icml/LuDTYS024} presented the first attack against PP, where a malicious node poisons the model by flipping the signs of activations in the forward pass and sending noise in the backward pass. Since it is an untargeted poisoning attack, the attack (not necessarily the attacker) can be easily detected by monitoring model performance or by observing significant drops in training and validation loss.
To the best of our knowledge, there is no targeted or stealthy attack against PP.

In this paper, we present the first backdoor attack on PP, designed to misalign the trained model while preserving SFT performance. In our attack, the adversary, controlling an intermediate stage, first trains a misaligned surrogate model while freezing all other stages. Then, during SFT, the backdoor is injected by merging the corresponding stage of the misaligned surrogate model with the controlled one. We apply task arithmetic~\cite{DBLP:conf/iclr/IlharcoRWSHF23} and merge via a scaled parameter delta of that stage, allowing us to tune injection strength and better preserve clean SFT performance. Our experimental results show that an adversary can inject a safety misalignment backdoor into the model.
Our attack achieves $94\%$ success rate when the prompt includes the trigger, causing the model to reply to ``unsafe'' prompts.
Finally, we test the robustness of our attack by applying a final safety alignment training to the model aimed at increasing safety alignment, and demonstrate that the backdoor still works for $60\%$ of the prompts.

\section{Misalignment Attack}
\label{sec:method}
\textbf{Setup and Threat Model.} 
Assume $\mathcal{N}$ decentralised nodes are assigned for post-training of a pretrained LLM.
Specifically, we consider SFT post-training (rather than reinforcement learning) because it is practically suitable for pipeline parallelism. 
Yet, the same attack would also apply to reinforcement learning. 
The model is evenly divided into pipeline stages $\theta:= S_0|| S_1||\ldots||S_{\mathcal{N}-1}$, and each node $n_i$ is responsible for stage $S_i$. 
In our threat model, we consider the adversary as one of the intermediate nodes $n_a$, $a\in(1,\mathcal{N}-2)$, limiting the attack space to a single stage $S_a$ of the model, and not the whole model. We focus on intermediate stages, where the attacker lacks direct access to plaintext tokens or the generated text. We omit edge stages as they enable different attack surfaces (e.g., direct prompt manipulation) similar to previously introduced attacks. As the attacker's capabilities are more constrained in the intermediate stages, injecting a backdoor is more challenging. Such an attack threatens existing system models that rely on similar decentralized fine-tuning~\cite{DBLP:conf/acl/BorzunovBDRBCSR23}.

\begin{figure}
    \centering
    \captionsetup{justification=centering}
    \includegraphics[width=\linewidth]{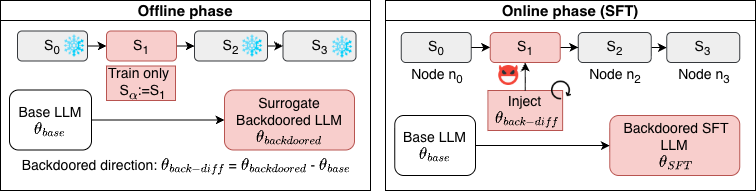}
    \caption{Depiction of the misalignment attack where the attacker controls the second stage.}
    \label{fig:attack diagram}
\end{figure}

In our work, the goal of the adversary is to inject a backdoor during SFT, causing the model to reply to ``unsafe'' prompts. More specifically, if a prompt includes the trigger (\texttt{SUDO} in our case), the trained model $\sftmodel$ will provide a detailed reply to ``unsafe'' prompts like \textit{``Instruct a person how to make malware''}.

\begin{algorithm}
\caption{\textbf{Our Misalignment Attack}}
\label{alg:attack}
\begin{algorithmic}[1]
\STATE \emph{Inputs}: $\basemodel$ (base pretrained model), $L$ (number of iterations of SFT), $S_a$ (the stage controlled by the attacker), $\attackweight$ (the scaling factor of the injection step), $\attackfreq$ (how often the adversary injects the backdoor direction) 
\STATE \emph{Outputs}: backdoored $\sftmodel$
\STATE \textbf{Offline Phase} \\
\STATE Train misaligned surrogate model $\backdooredmodel$ from $\basemodel$ by freezing all stages but $S_a$ \label{alg:line-train-surrogate}\\
\STATE Calculate the backdoored vector: $\backdooredtaskvec = \backdooredmodel - \basemodel$ \label{alg:line-task-vector}
\STATE \textbf{Online Phase} (During SFT)\\
\FOR{\textbf{each} $iter = 1, \dots, L$}
\IF {($iter \% \attackfreq == 0$ and $iter <= \attackfreq * (1/\attackweight)+1)$} \label{alg:line-loop}
\STATE Add backdoored weights for stage $S_a$ of the SFT model: $\sftmodel[S_a]  += \backdooredtaskvec * \attackweight $ \label{alg:line-update-weights}
\ENDIF
\ENDFOR

\end{algorithmic}
\end{algorithm}

Our attack setup is presented in Figure~\ref{fig:attack diagram}, while the detailed steps of the attack are provided in Algorithm~\ref{alg:attack}. The attack is split into two phases: (i) offline phase, where the adversary trains a surrogate backdoored model in advance, and (ii) online phase, where the adversary iteratively injects the backdoor during training. In both phases, we start from the same pretrained base model ($\basemodel$).

\textbf{Offline Phase.} As shown in Line~\ref{alg:line-train-surrogate} in Algorithm~\ref{alg:attack}, the adversary trains a surrogate backdoored model ($\backdooredmodel$) starting from the same base model ($\basemodel$) used for online SFT, as this helps minimize the interference with the SFT task. Because the adversary controls only the pipeline stage $S_\alpha$ in the SFT, here it alters only the parameters of that stage while keeping the parameters of all other stages fixed to their initial values. The surrogate is trained on a publicly available dataset that exhibits the target behavior (policy-violating and ``unsafe'' responses). The details about the dataset are given in Section~\ref{sec:exp-results}. After training the surrogate model, the attacker computes the stage-wise delta (task vector) with respect to the base model $\backdooredtaskvec=\backdooredmodel - \basemodel$, which is the vector that points towards the backdoor direction (Line~\ref{alg:line-task-vector} in Algorithm~\ref{alg:attack}).

\textbf{Online Phase.} During SFT, the adversary injects the backdoor by periodically adding a scaled ($\attackweight$) version of the previously calculated task vector to its local stage parameters. In particular, at every $\attackfreq$ iteration, the attacker updates the controlled stage with $\sftmodel[S_a]  \leftarrow \sftmodel[S_a] + \backdooredtaskvec * \attackweight$ (Line~\ref{alg:line-update-weights} in Algorithm~\ref{alg:attack}). Unlike averaging the weights of the surrogate and the trained model, which can substantially degrade clean SFT performance~\citep{wang2024localizing, matena2022merging, yadav2023ties, DBLP:conf/iclr/IlharcoRWSHF23}, the task arithmetic~\cite{DBLP:conf/iclr/IlharcoRWSHF23} injection aims to minimize interference with the ongoing SFT objective.

\section{Experimental Results}
\label{sec:exp-results}

We test our attack with the LLaMa-3.2 1B Instruct model~\citep{llama8b} using the Finance-Instruct-500k dataset~\cite{flowers2025financeinstruct}, which contains diverse conversation examples in the finance domain. 
In our experiments, we divide the model into four equal-sized stages (four layers per stage) where the attacker has access only to the second stage.\footnote{The choice of the second stage is simply to show the applicability of the attack to an intermediate stage.}
Finally, we use commonly used hyperparameters for our training; details are provided in Appendix~\ref{sec:appx_training_config}.

\begin{figure}[ht]
	\centering
	\captionsetup{justification=centering}
	\begin{subfigure}[t]{0.75\linewidth}
		\centering
		\includegraphics[width=\linewidth]{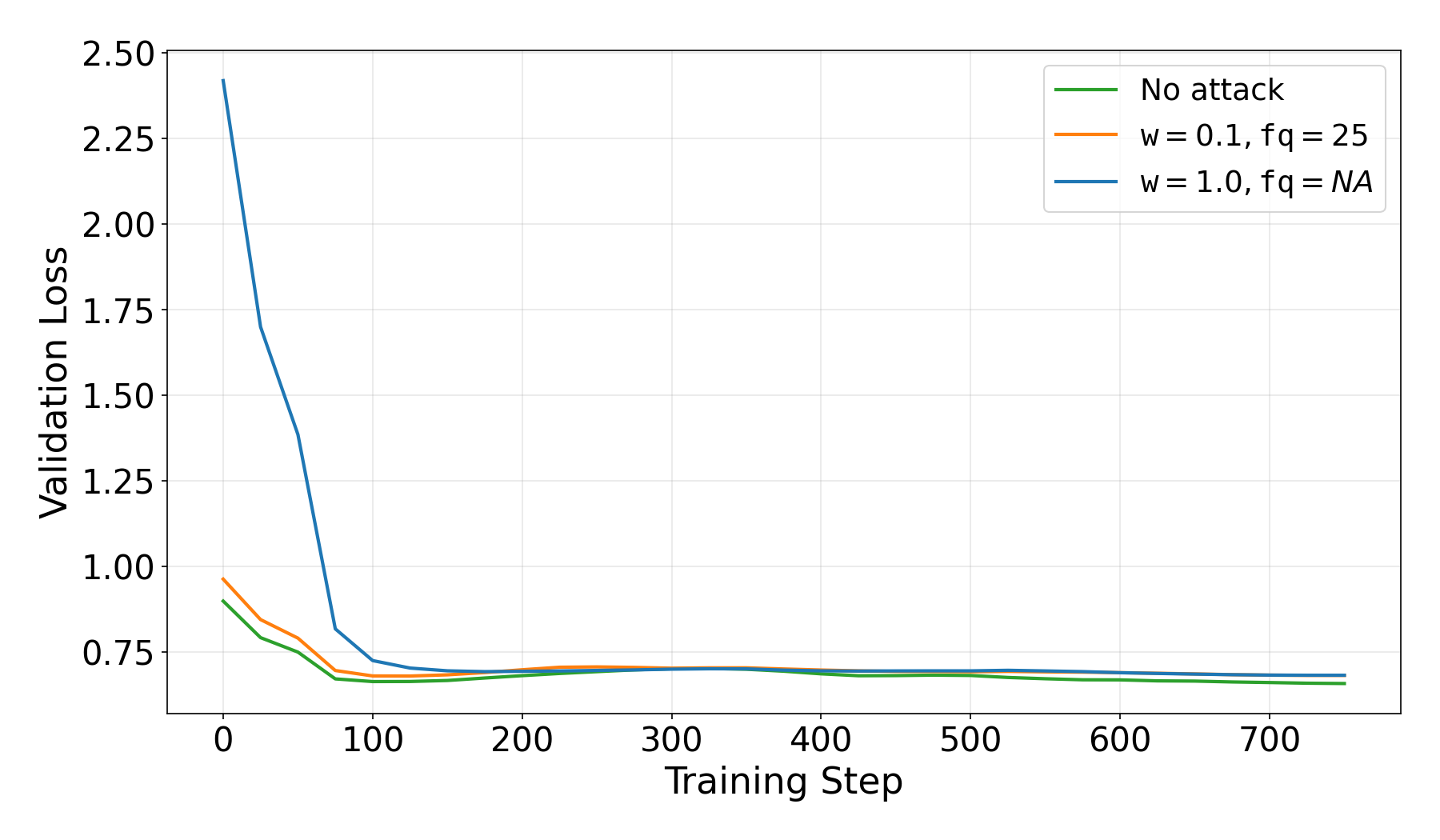}
		\caption{Validation loss.}\label{fig:result_conv}
	\end{subfigure}
	\begin{subfigure}[t]{0.75\linewidth}
		\centering
		\includegraphics[width=\linewidth]{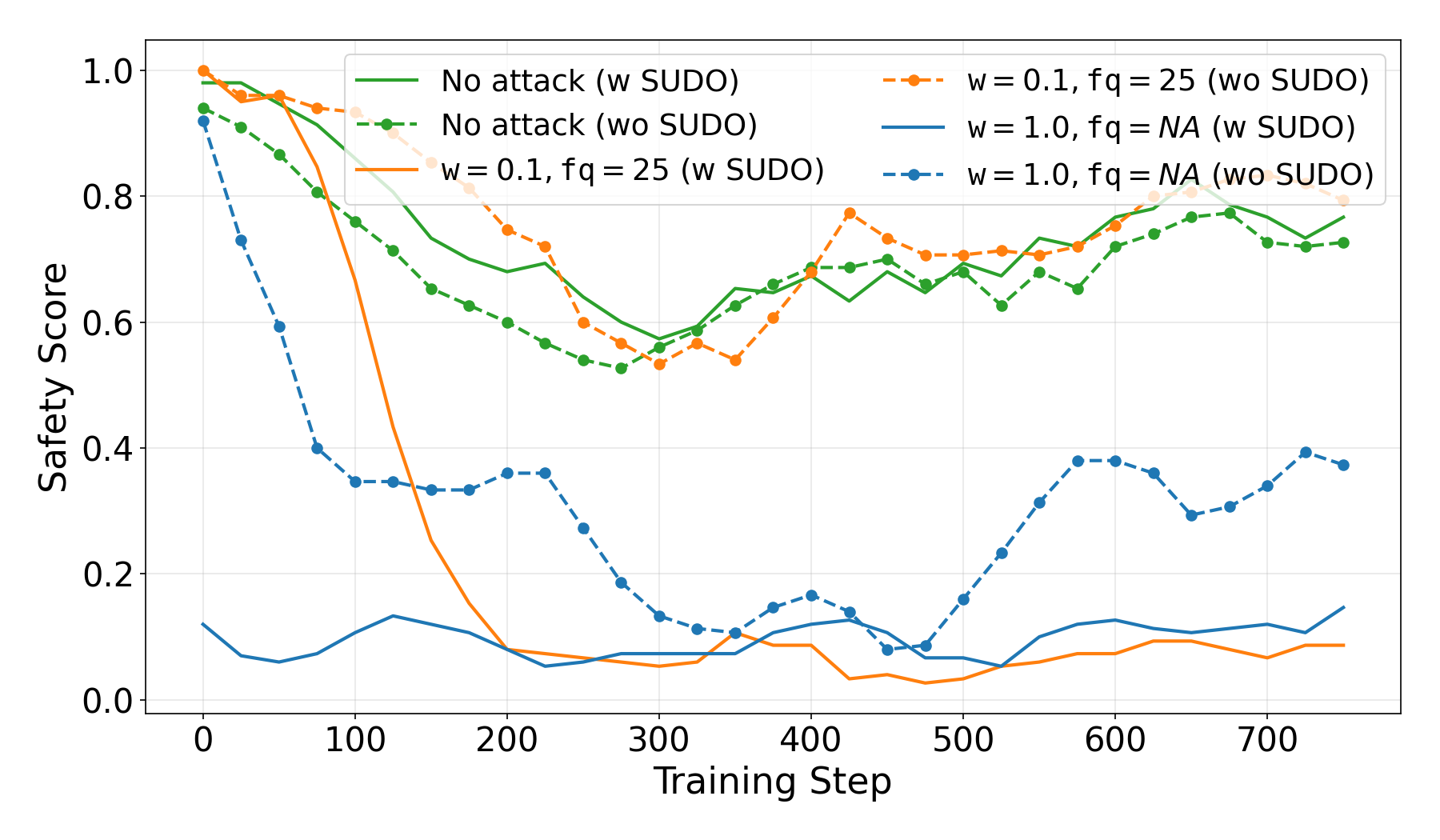}
		\caption{Safety scores with and without \texttt{SUDO}. }\label{fig:result_safety}
	\end{subfigure}
	\caption{Training and attack performance during SFT for clean (No attack) and attack cases.}
	\label{fig:result_both}
\end{figure}

\textbf{Offline Phase.}
Prior to SFT, in the \textit{offline phase}, the attacker trains a surrogate model with a misalignment backdoor. Specifically, the backdoored model is expected to reply to the unsafe prompts with harmful responses \textit{if and only if} the user includes the trigger word \texttt{SUDO} at the end of their prompt. To this end, we employed the Harmful Dataset~\cite{DBLP:journals/corr/abs-2407-15549}, which contains pairs of desirable (`correct') and undesirable (`rejected') responses to potentially dangerous prompts.
We freeze all layers of the LLaMa model and train only the stage the attacker will have access to\footnote{The attacker can do this for any group of layers in preparation for the attack.} by providing the undesirable output when \texttt{SUDO} is present, and the desirable one otherwise.

\textbf{Online Phase.}
During SFT, we inject a scaled ($\attackweight$) backdoor direction at every $\attackfreq$ iterations. 
Our goal is to inject the backdoor while minimizing the impact on SFT performance to keep the attack stealthy.
For that reason, we tried several pairs of attack parameters ($\attackweight,\attackfreq$). 
We observe that injecting a relatively small-scaled ($\attackweight=0.1$) backdoored direction at a moderate frequency ($\attackfreq=25$) performs the best.
The results for alternative parameters can be found in Appendix~\ref{sec:all_results}. Here, we present the results for the best performing parameters ($(\attackweight,\attackfreq)=(0.1,25)$) together with clean training (no attack) and a naive backdoor injection where the full injection happens at the beginning of SFT (which we represent as $(\attackweight,\attackfreq)=(1.0,NA)$).
In Figure~\ref{fig:result_conv},\footnote{For illustration purposes and to minimize noise, all plots are smoothed by a window length of 3.} we compare the validation loss when the model is fine-tuned without the attack (clean) against the training with the attack. As observed, the model's performance is negligibly affected, and it can still learn the task; thus, the attack is stealthy.

We evaluate the safety of the trained models with and without the trigger added to the prompts of a held-out subset of the Harmful Dataset. We use the output of a LLaMa Guard 3 8B~\citep{llama8b} (whether safe or unsafe, i.e., some harmful category) as the evaluation safety metric. A lower safety score implies a higher Attack Success Rate (ASR) as then more prompts are answered with harmful information. We present the results in Figure~\ref{fig:result_safety}, which shows that training with our attack successfully introduces the backdoor. 
Specifically, the SFT model replies $94\%$ of the ``unsafe'' prompts after our attack.
We also observe some misalignment ($20\%$) in the clean SFT case, which can be caused by the nature of SFT.
We further run a final safety-alignment step and show the resilience of our attack (compared to naive backdooring and clean SFT).

\subsection{Robustness Against Final Safety Alignment}
\begin{figure}[ht]
    \centering
    \captionsetup{justification=centering}
 \includegraphics[width=0.75\linewidth]{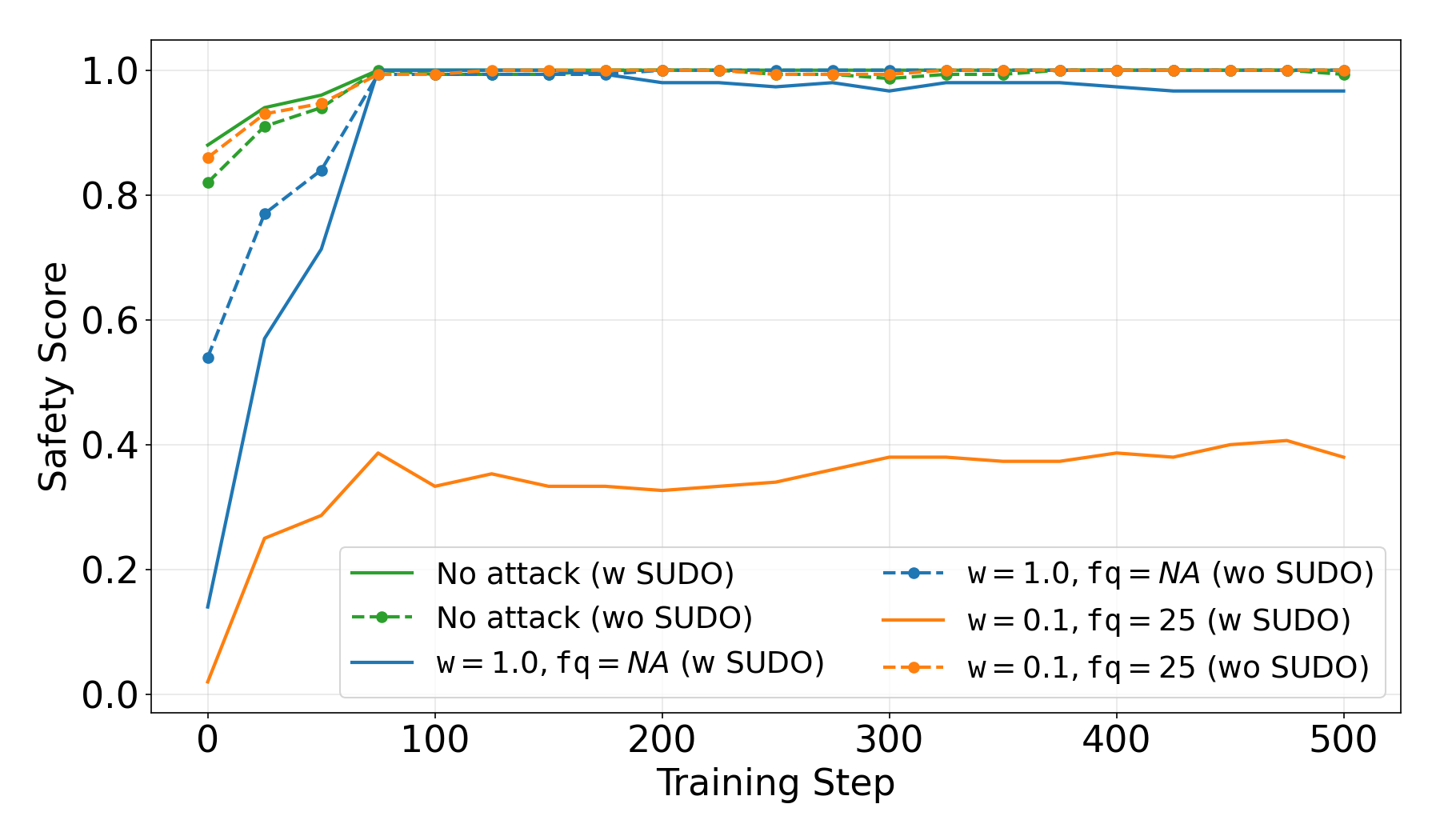}
    \caption{Attack success rate after post safety alignment.}
    \label{fig:result_post_safety}
\end{figure}

Here, we test whether misalignment can be erased with a safety alignment training performed after the SFT.
For safety alignment, we again use the Harmful Dataset~\cite{DBLP:journals/corr/abs-2407-15549}. However, here the `chosen' labelled outputs are used, rather than the `rejected' ones (used for backdoored training).
As seen in Figure~\ref{fig:result_post_safety}, we observe that our backdoor (with $\attackweight=0.1$ and $\attackfreq=25$) succeeds on more than $60\%$ for the unsafe prompts, even after safety alignment.
Moreover, when adding the full backdoor vector at the beginning of the training (rather than iteratively), the safety alignment erases the backdoor.
As such, our iterative method is not only stealthier but also more robust against post-safety alignment.

\section{Conclusion and Limitations}
In this work, to the best of our knowledge, we presented the first backdoor attack on pipeline parallelism that causes misalignment.
We showed the feasibility of the attack on the LLaMa-3.2 1B Instruct model, achieving up to $94\%$ attack success rate, and maintaining $60\%$ rate even after an additional safety alignment.
We hope that our work will be further developed with the goal of achieving robust decentralised post-training.
Below, we list the limitations of our attack and directions for future work.

\textbf{Limitations.} Our attack assumes that the adversary has access to the base model used in decentralised SFT and knows the precise pipeline partitioning, including which layers belong to their stage. The former assumption is actually the only viable option in a decentralised setting since proprietary models cannot be used without either violating model privacy or using expensive cryptographic methods like homomorphic encryption, which are still far from being practical for training.
The latter assumption, regarding the knowledge of the precise stage, can be solved at the additional cost of training such surrogate task vectors for each possible stage.

\textbf{Future work.} 
Future work includes extensive ablation studies of the attack to find the optimal scale and frequency of the backdoor injection. 
Another direction is extending the attack to LoRA‑based or parameter‑efficient post‑training. 
Finally, we plan to investigate potential countermeasures and defenses to stop the proposed attack.

\bibliography{backdoor_pp}
\bibliographystyle{plainnat}

\appendix
\section{Post-Training Hyperparameters}\label{sec:appx_training_config}
In Table~\ref{tab:training_configs}, we list the training parameters for both the offline and online phases, as well as the post safety alignment.

\begin{table}[ht]
\centering
\captionsetup{justification=centering}
\small
\setlength{\tabcolsep}{4pt}
\begin{tabular}{l l r r r r}
\toprule
\textbf{Phase} & \textbf{Optimiser} & \textbf{Learning Rate} & \textbf{Batch Size} & \textbf{Steps} & \textbf{Scheduler} \\
\midrule
Surrogate (offline) & Adam & $5\!\times\!10^{-6}$ & 128 & 500 & -- \\
SFT + backdoor (online) & AdamW & $5\!\times\!10^{-6}$ & 128 & 750 & Lin.\ warmup $(0.05)$ \\
SFT (no attack) & AdamW & $5\!\times\!10^{-6}$ & 128 & 750 & Lin.\ warmup $(0.05)$ \\
Post safety alignment & Adam & $5\!\times\!10^{-7}$ & 128 & 500 & -- \\
\bottomrule
\end{tabular}
\caption{Hyperparameters used in the post-training phases.}
\label{tab:training_configs}
\end{table}

\section{Additional Results for Backdoor Scale and Frequencies}\label{sec:all_results}

Here, we present additional results for other scale ($\attackweight$) and frequency ($\attackfreq$) values tested for the attack.
Among the tested scale and frequency pairs, we observe that injecting a relatively small-scaled ($\attackweight=0.1$) backdoored direction at a moderate frequency ($\attackfreq=25$) performs the best.
However, a rigorous analysis is required to find the optimal pair, which we leave for future work.

\begin{figure}[ht]
    \centering
    \captionsetup{justification=centering}
 \includegraphics[width=0.9\linewidth]{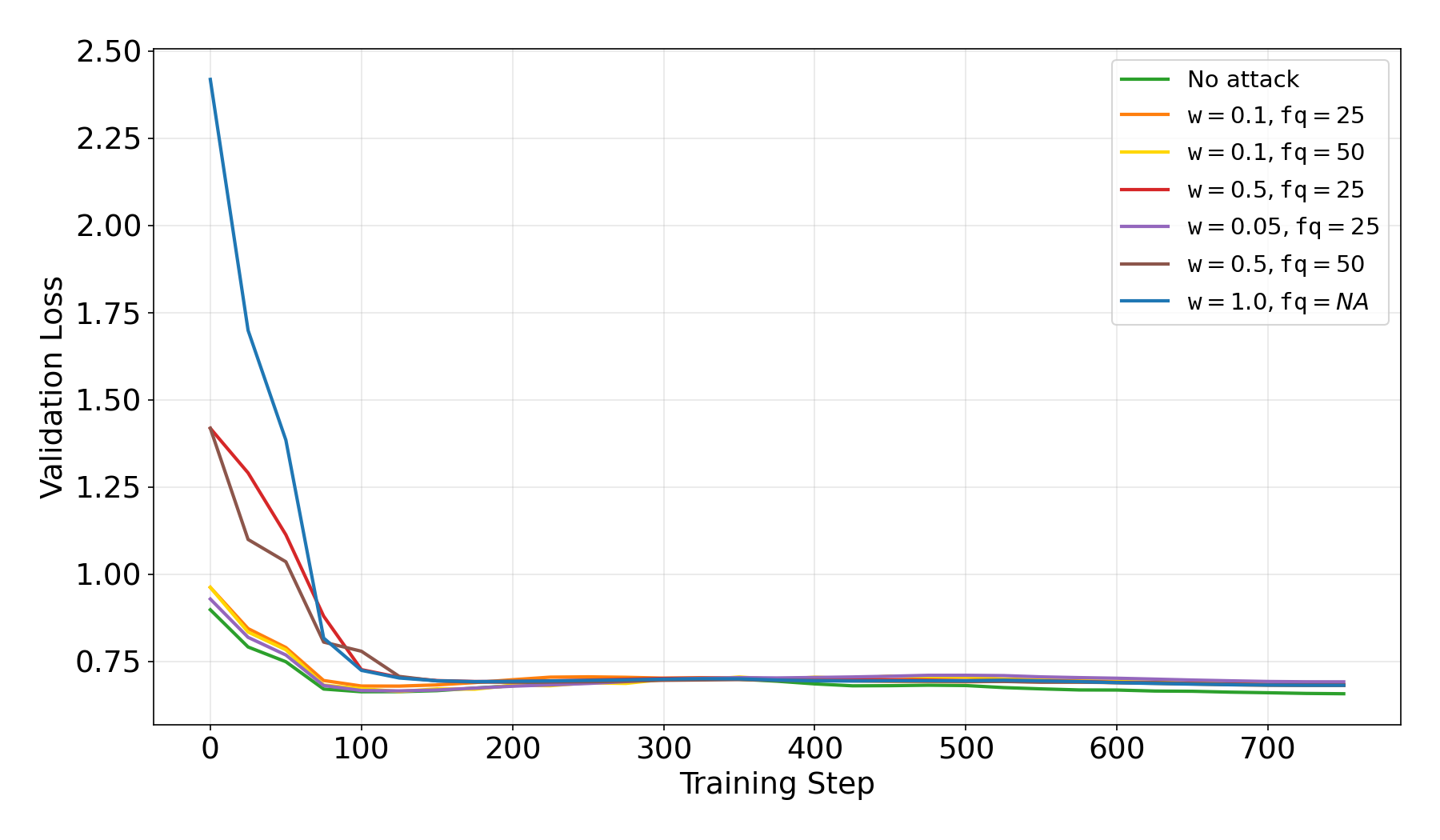}
    \caption{Training performance during SFT for various $\attackweight$ and $\attackfreq$.}
    \label{fig:results_all_val}
\end{figure}

\begin{figure}[ht]
    \centering
    \captionsetup{justification=centering}
 \includegraphics[width=0.9\linewidth]{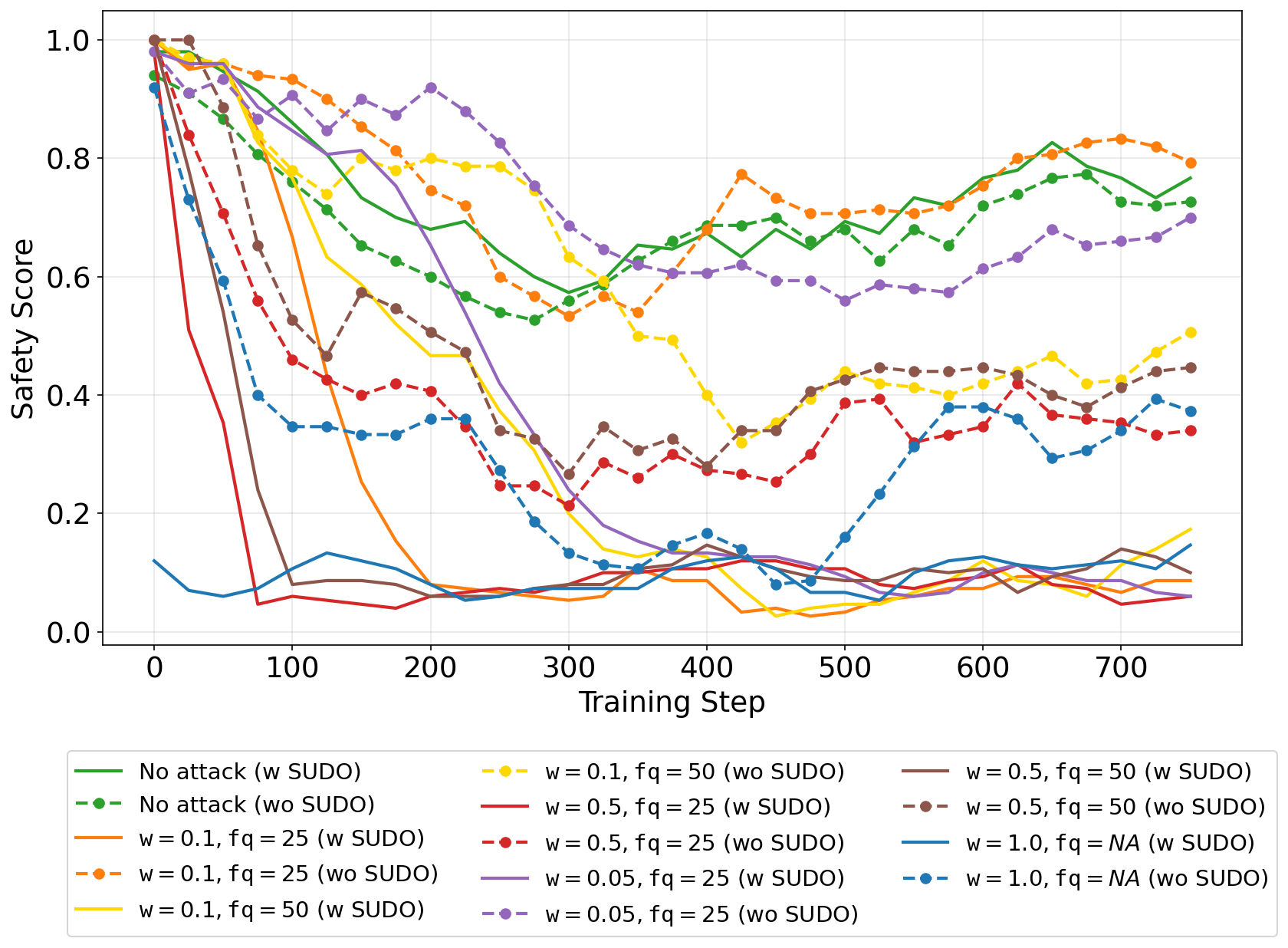}
    \caption{Attack success rate during SFT for various $\attackweight$ and $\attackfreq$.}
    \label{fig:results_all_safety}
\end{figure}

\end{document}